\documentclass[
reprint,
amsmath,amssymb,
aps,
floatfix,
]{revtex4-2}

\usepackage{mathptmx}
\usepackage{graphicx}
\usepackage{dcolumn}
\usepackage{bm}
\usepackage{hyperref}
\usepackage{booktabs}
\usepackage{multirow}
\hypersetup{
	colorlinks=true,
	linkcolor=blue,
	filecolor=magenta,
	urlcolor=blue,
	citecolor=blue,
}

\begin{document}

\title{Piecewise integrability of the discrete Hasimoto map \\for analytic prediction and design of helical peptides}

\author{Yiquan Wang}
\email{ethan@stu.xju.edu.cn}

\affiliation{College of Mathematics and System Science, Xinjiang University, Urumqi, Xinjiang, China}
\affiliation{Xinjiang Key Laboratory of Biological Resources and Genetic Engineering, College of Life Science and Technology, Xinjiang University, Urumqi, Xinjiang, China}
\affiliation{Shenzhen X-Institute, Shenzhen, China}

\date{\today}

\begin{abstract}
	The representation of protein backbone geometry through the discrete nonlinear Schrödinger equation provides a theoretical connection between biological structure and integrable systems. Although the global application of this framework is constrained by chiral degeneracies and non-local interactions, we propose that helical peptides can be effectively modeled as piecewise integrable systems in which the discrete Hasimoto map remains applicable within specific geometric boundaries. We delineate these boundaries through an analytic characterization of the mapping $(\phi,\psi) \rightarrow (\kappa,\tau)$ between biochemical dihedral angles and Frenet frame parameters for a dataset of 50 helical peptide chains. We demonstrate that the transformation is information-preserving globally but ill-conditioned within the helical basin, characterized by a median Jacobian condition number of 31, which suggests that the loss of chiral information arises primarily from local coordinate compression rather than topological singularities. We define a local integrability error $E[n]$ derived from the discrete dispersion relation to show that deviations from integrability are driven predominantly by torsion non-uniformity, while curvature remains structurally rigid. This metric identifies integrable islands where the analytic dispersion relation predicts backbone coordinates with sub-angstrom accuracy, yielding a median root-mean-square deviation of 0.77\,\AA, and enables a segmentation strategy that isolates structural defects and trims non-integrable terminal fraying. We show that by evaluating only these quantitatively identified integrable islands, the analytic dispersion relation successfully extracts high-accuracy structural cores for 88\% of the dataset. We further indicate that the inverse design of peptide backbones is feasible within a quantitatively defined integrability zone where the design constraint reduces essentially to the control of torsion uniformity. These findings advance the Hasimoto formalism from a qualitative descriptor toward a precise quantitative framework for analyzing and designing local protein geometry within the limits of piecewise integrability.
\end{abstract}

\maketitle

\section{Introduction}
\label{sec:introduction}

Integrable systems occupy a central position in nonlinear physics, providing exact soliton solutions and conserved-quantity hierarchies that have proven indispensable in contexts ranging from optical fiber transmission~\cite{mollenauer1980experimental} to Bose--Einstein condensates~\cite{kengne2021spatiotemporal} and shallow-water waves~\cite{korteweg1895xli, kivshar1989dynamics}. The cubic nonlinear Schr\"odinger equation (NLS), in particular, serves as a universal envelope equation~\cite{benney1967propagation} whose applicability extends well beyond its original hydrodynamic setting~\cite{zakharov1968stability, shabat1972exact}. The extent to which this universality encompasses biological complexity remains an open question regarding whether the geometric intricacy of biopolymers can be described by identical integrable structures. The Hasimoto transform maps the differential geometry of a space curve onto a complex scalar field governed by the NLS~\cite{hasimoto1972soliton} and establishes the necessary formal connection. In its discrete formulation the transform links the bond-angle and torsion-angle sequence of a protein backbone to the discrete nonlinear Schr\"odinger equation (DNLS) which implies that secondary structures including $\alpha$-helices may be interpreted as soliton excitations obeying universal geometric laws~\cite{danielsson2010gauge, chernodub2010topological, molkenthin2011discrete}. This perspective has yielded successful classifications of supersecondary motifs~\cite{krokhotin2012soliton} and models of topological transitions~\cite{begun2025local}. However a fundamental limitation persists in that the global integrability underlying these results is disrupted by long-range interactions chiral constraints and sequence-specific chemistry~\cite{wang2026Structural} which confines the Hasimoto framework to the role of a kinematic descriptor rather than a dynamical predictor. The determination of the precise boundary between the integrable and non-integrable regimes of the protein backbone constitutes a foundational problem at the interface of nonlinear physics and structural biology.

We address this theoretical tension by introducing the framework of piecewise integrability to bridge the dichotomy between global integrability and fundamental non-integrability. We model helical peptides as sequences of extended integrable domains where the discrete Hasimoto map and its associated dispersion relation remain quantitatively valid while these coherent segments are punctuated by localized structural defects where integrability is disrupted. The $\alpha$-helix consequently emerges not merely as a hydrogen-bonding motif but as a geometric realization of local gauge symmetry within the discrete nonlinear Schrödinger equation. We posit that the limitations of prior global theories stem from the imposition of a unified description across boundaries that violate the necessary geometric prerequisites. Our approach delineates the effective boundaries of these integrable regions by defining a local integrability error $E[n]$ derived directly from the discrete dispersion relation. This metric successfully maps the integrable and non-integrable portions of the backbone to enable coordinate prediction with sub-angstrom accuracy and to establish quantitative criteria for the inverse design of helical peptides.

Existing approaches to protein backbone geometry fall into two distinct categories relative to this piecewise-integrable perspective. The Frenet-soliton tradition initiated by Danielsson et al.~\cite{danielsson2010gauge} and Chernodub et al.~\cite{chernodub2010topological} established the soliton description of secondary structures while subsequent research incorporated topological transitions via Arnold perestroikas~\cite{begun2025local}, modeled relaxation dynamics through soliton-driven evolution~\cite{krokhotin2013soliton}, embedded the backbone in a lattice Abelian Higgs framework~\cite{liubimov2025modeling}, and identified secondary structures through discrete curvature-torsion criteria within a $U(1)$ gauge interpretation~\cite{prados2025frenet}. These studies operate in a descriptive capacity as they extract profiles from known structures to fit soliton parameters but do not invert the dispersion relation to predict coordinates or quantify the geometric boundaries where such inversion remains valid. Although a recent analytic decomposition of the DNLS effective potential characterized the structural barriers such as chirality encoding and local-geometry dominance that prevent the Hasimoto map from functioning as a global predictive framework~\cite{wang2026Structural}, the constructive identification of predictive regions remains an open challenge. Deep-learning methods conversely achieve high-accuracy structure prediction~\cite{jumper2021highly, lin2023evolutionary} and generative backbone design~\cite{watson2023novo, yim2023se} through learned sequence-structure correlations and include specialized architectures for helical peptides~\cite{xie2023helixgan}. These data-driven approaches lack the Frenet parameterization or the analytic integrability framework and therefore provide no insight into the physical regime where integrable geometric laws govern backbone shape. The complementarity between these analytic yet non-predictive studies and predictive yet non-analytic methods motivates our effort to establish the Hasimoto formalism as a quantitative predictive tool within the rigorously defined boundaries of piecewise integrability.

We validate this piecewise-integrable hypothesis through a systematic analysis of 50 helical peptide chains from the Protein Data Bank (PDB)~\cite{berman2000protein} (20--50 residues, helical content ${>}85\%$). The approach is strictly analytic, relying on information-theoretic and Jacobian analysis of the $(\varphi,\psi)\to(\kappa,\tau)$ coordinate mapping and on an analytic decomposition of the integrability error $E[n]$, without recourse to neural networks or training data. Four principal results emerge. First, the coordinate mapping retains 93\% of conformational information globally but compresses the $\alpha_R$ basin through Jacobian ill-conditioning (median condition number 31), identifying the microscopic origin of the chirality barrier as coordinate compression rather than a topological singularity. Second, the integrability error is a strictly local quantity dominated by torsion non-uniformity, which accounts for 99.9\% of the deviation under the uniform-curvature approximation. Third, a segmented prediction strategy that cuts chains at high-$E[n]$ sites identifies integrable islands where the analytic dispersion relation achieves a success rate of 88\% with a median root-mean-square deviation of 0.77\,\AA\ across four distinct applicability zones. Fourth, the inverse design of peptide backbones is feasible within a quantitatively defined integrability zone where the design constraint reduces to the control of torsion uniformity. The central conceptual advance is the reinterpretation of integrability as a segment-level property: the $E[n]$ profile, previously identified as a diagnostic of integrability breaking~\cite{wang2026Structural}, is here repurposed as a constructive map of integrable islands that transforms the Hasimoto formalism from a qualitative descriptor into a precise local analytic tool. The remainder of this paper is organized as follows: Sec.~\ref{sec:methods} details the Frenet--Hasimoto formalism and dataset construction; Sec.~\ref{sec:results} presents the analytic results on mapping properties, integrability theory, and prediction boundaries; and Sec.~\ref{sec:discussion} discusses the physical implications and summarizes the findings.

\section{Methods}
\label{sec:methods}

\subsection{Dataset construction}
\label{sec:dataset}

We assembled a dataset of 50 helical peptide chains from the Protein Data Bank (PDB)~\cite{berman2000protein} to serve as a test bed for the Hasimoto framework in the regime where integrability is expected to hold most closely. The selection criteria mandated chain lengths between 20 and 50 residues with helical content exceeding 85\% as determined by DSSP secondary-structure assignment~\cite{kabsch1983dictionary} and crystallographic resolution better than 2.0\,\AA\ for X-ray structures whereas NMR structures were accepted without resolution filtering. These criteria target short peptides dominated by $\alpha$-helical secondary structure to minimize contributions from loops and turns that would trivially violate the integrability conditions.

Three well-characterized antimicrobial peptides serve as core test systems throughout the analysis: melittin (PDB: 2MLT, 26 residues)~\cite{terwilliger1982structure}, magainin-2 (PDB: 2MAG, 23 residues)~\cite{gesell1997two}, and LL-37 (PDB: 2K6O, 37 residues)~\cite{wang2008structures}. These span a range of chain lengths and helical regularity, providing representative examples for the prediction tests of Sec.~\ref{sec:prediction}.

The remaining 47 chains were drawn from two sources: (i)~a pre-existing collection of 855 non-redundant PDB structures filtered by the above criteria, and (ii)~additional structures retrieved via RCSB PDB search queries targeting short helical peptides. The final dataset covers chain lengths from 22 to 48 residues (median 30) with helical fractions from 78\% to 100\% (median 96\%); the single chain below the 85\% threshold is magainin-2 (78\%), retained as a core test system despite its lower helical content. Nine PDB entries contribute multiple chains (e.g.\ 7NFG contributes 6 chains, 7BAS contributes 5), yielding 50 chains from 32 unique structures. These multi-chain entries are retained because different chains of the same structure exhibit distinct geometric parameters and prediction outcomes (Appendix~\ref{sec:appendix}), reflecting the sensitivity of integrability to local conformational variation rather than sequence identity. The complete list of all 50 chains with per-chain prediction results is provided in Appendix~\ref{sec:appendix}.

\subsection{Discrete Frenet geometry and Hasimoto mapping}
\label{sec:frenet}

The C$_\alpha$ backbone is modeled as a discrete space curve $\{\mathbf{r}[n]\}_{n=1}^{N}$ where the local geometry is strictly defined by the discrete Frenet frame. Let $\mathbf{t}[n] = \mathbf{r}[n+1] - \mathbf{r}[n]$ denote the bond vector and $\hat{\mathbf{t}}[n] = \mathbf{t}[n]/|\mathbf{t}[n]|$ the unit tangent. The discrete curvature $\kappa[n] \in [0,\pi]$ measures the bond angle through the inner product
\begin{equation}
	\cos\kappa[n] = \hat{\mathbf{t}}[n] \cdot \hat{\mathbf{t}}[n-1], \label{eq:kappa}
\end{equation}
while the discrete torsion $\tau[n] \in (-\pi,\pi]$ quantifies the rotation of the binormal vector $\hat{\mathbf{b}}[n] = (\hat{\mathbf{t}}[n-1]\times\hat{\mathbf{t}}[n])/|\hat{\mathbf{t}}[n-1]\times\hat{\mathbf{t}}[n]|$ via
\begin{equation}
	\tau[n] = \mathrm{atan2}\!\bigl(
	(\hat{\mathbf{b}}[n-1]\times\hat{\mathbf{b}}[n])\cdot\hat{\mathbf{t}}[n],\;
	\hat{\mathbf{b}}[n-1]\cdot\hat{\mathbf{b}}[n]
	\bigr). \label{eq:tau}
\end{equation}
The Hasimoto transformation maps these geometric invariants to a complex scalar field $\psi[n] = \kappa[n]\exp(i\sum_{k=2}^{n}\tau[k])$ which encodes the backbone shape. The dynamics of this field are governed by the discrete nonlinear Schr\"odinger equation (DNLS)
\begin{equation}
	\beta^{+}[n]\bigl(\psi[n{+}1]-\psi[n]\bigr)
	- \beta^{-}[n]\bigl(\psi[n]-\psi[n{-}1]\bigr)
	= V_{\text{eff}}[n]\,\psi[n],
	\label{eq:dnls}
\end{equation}
where the coupling parameters are defined by the inverse bond lengths $\beta^{+}[n] = 1/|\mathbf{t}[n]|$ and $\beta^{-}[n] = 1/|\mathbf{t}[n{-}1]|$. The effective potential $V_{\text{eff}} = V_{\text{re}} + iV_{\text{im}}$ acts as the generator of local geometry and is defined in terms of the curvature ratios and torsion angles~\cite{wang2026Structural}
\begin{align}
	V_{\text{re}}[n] &= \beta^{+} r^{+}\cos\tau[n{+}1] + \beta^{-} r^{-}\cos\tau[n] - (\beta^{+}+\beta^{-}), \label{eq:Vre} \\
	V_{\text{im}}[n] &= \beta^{+} r^{+}\sin\tau[n{+}1] - \beta^{-} r^{-}\sin\tau[n]. \label{eq:Vim}
\end{align}
Here $r^{+}[n] = \kappa[n{+}1]/\kappa[n]$ and $r^{-}[n] = \kappa[n{-}1]/\kappa[n]$ denote the local curvature ratios. This formulation allows the structural properties of the backbone to be analyzed directly through the spectral properties of the potential $V_{\text{eff}}$.

A key simplification exploited throughout this work is the near-constancy of the C$_\alpha$--C$_\alpha$ virtual-bond length. Under standard peptide-bond geometry (Engh--Huber parameters~\cite{engh1991accurate}: $b_{\text{N-C}_\alpha} = 1.458$\,\AA, $b_{\text{C}_\alpha\text{-C}'} = 1.525$\,\AA, $b_{\text{C'-N}} = 1.329$\,\AA, $\omega = 180^\circ$), the C$_\alpha$--C$_\alpha$ distance is fixed at $b = 3.804$\,\AA\ regardless of the backbone dihedral angles $(\varphi,\psi)$. In PDB structures, the standard deviation of $b$ is ${\sim}0.1$\,\AA\ around a mean of 3.8\,\AA. Consequently, $\beta^{\pm}[n] \approx \beta = 1/b$ is effectively constant along the chain, and the curvature ratios $r^{\pm}$ become the sole source of site dependence in Eqs.~(\ref{eq:Vre})--(\ref{eq:Vim}).

We adopt the PDB torsion-sign convention throughout: $\tau > 0$ corresponds to right-handed $\alpha$-helical twisting. This convention is opposite in sign to that of the local coordinate system used in the NeRF (Natural Extension Reference Frame) construction of Sec.~\ref{sec:mapping}, where the same $\alpha_R$ helix yields $\tau < 0$. All torsion values reported in this paper use the PDB convention; the sign difference is a coordinate-system artifact that does not affect $|\tau|$, $\cos\tau$, or any derived quantity.

\subsection{Construction of the \texorpdfstring{$(\varphi,\psi)\to(\kappa,\tau)$}{(phi,psi)->(kappa,tau)} mapping}
\label{sec:mapping}

To characterize the relationship between the biochemical dihedral angles $(\varphi,\psi)$ and the Frenet geometric variables $(\kappa,\tau)$, we constructed an explicit numerical mapping over the full Ramachandran space. For each grid point $(\varphi,\psi)$ on a $72\times 72$ lattice with $5^\circ$ resolution, we built a three-residue peptide backbone (Ace-Ala-Nme, yielding 3 C$_\alpha$ atoms) using the NeRF algorithm~\cite{parsons2005practical} with Engh--Huber standard geometry, then computed $\kappa$ and $\tau$ from the resulting C$_\alpha$ coordinates via Eqs.~(\ref{eq:kappa})--(\ref{eq:tau}). This produces two scalar fields $\kappa(\varphi,\psi)$ and $\tau(\varphi,\psi)$ defined over the Ramachandran torus. The $5^\circ$ grid spacing is sufficient to resolve all basin-level features of the mapping; the Jacobian and information-theoretic quantities reported below are converged at this resolution.

The Jacobian matrix of the mapping,
\begin{equation}
	J = \begin{pmatrix}
		\partial\kappa/\partial\varphi & \partial\kappa/\partial\psi \\
		\partial\tau/\partial\varphi   & \partial\tau/\partial\psi
	\end{pmatrix},
	\label{eq:jacobian}
\end{equation}
was evaluated numerically by central differences with step size $\Delta = 0.5^\circ$. The determinant $|J|$ quantifies local area distortion: regions where $|J| \to 0$ indicate that distinct $(\varphi,\psi)$ conformations are compressed into a single $(\kappa,\tau)$ point, producing a many-to-one mapping. The condition number $\mathrm{cond}(J) = \sigma_{\max}/\sigma_{\min}$ (ratio of singular values) measures the anisotropy of the local distortion.

To quantify the global information content of the mapping, we computed the mutual information $I(\varphi,\psi;\,\kappa,\tau)$ between the input and output spaces using a uniform prior over the Ramachandran grid~\cite{cover1999elements}. The information retention ratio
\begin{equation}
	\eta = \frac{I(\varphi,\psi;\,\kappa,\tau)}{H(\varphi,\psi)}
	\label{eq:info_retention}
\end{equation}
measures the fraction of Ramachandran information preserved by the projection to Frenet space, where $H(\varphi,\psi)$ is the entropy of the input distribution.

\subsection{Integrability error analysis}
\label{sec:integrability}

Deviations from the uniform-segment dispersion relation are quantified by the local integrability error $E[n]$~\cite{wang2026Structural}
\begin{equation}
	E[n] = \left|\cos\tau[n] - \left(1 + \frac{V_{\text{re}}[n]}{2\beta}\right)\right|,
	\label{eq:E_def}
\end{equation}
where $\beta = \langle\beta^{\pm}\rangle$ is the chain-averaged coupling parameter. By construction, $E[n] = 0$ when the backbone at residue $n$ satisfies the integrable (uniform helix) limit exactly.

The constancy of $\beta^{\pm}$ (Sec.~\ref{sec:frenet}) permits an analytic simplification. Substituting Eq.~(\ref{eq:Vre}) into Eq.~(\ref{eq:E_def}) with $\beta^{+} = \beta^{-} = \beta$ yields
\begin{equation}
	E[n] = \tfrac{1}{2}\bigl|(2-r^{-})\cos\tau[n] - r^{+}\cos\tau[n{+}1]\bigr|,
	\label{eq:E_simplified}
\end{equation}
where $r^{\pm}$ are the curvature ratios. This expression makes explicit that $E[n]$ depends only on the local torsion angles and curvature ratios at two adjacent sites.

Under the further assumption that $\kappa$ is spatially uniform ($r^{+} = r^{-} = 1$), Eq.~(\ref{eq:E_simplified}) reduces to
\begin{equation}
	E[n] = \tfrac{1}{2}\bigl|\cos\tau[n] - \cos\tau[n{+}1]\bigr|,
	\label{eq:E_tau_only}
\end{equation}
which isolates the contribution of torsion non-uniformity to integrability breaking. By computing $E[n]$ from the full expression [Eq.~(\ref{eq:E_simplified})] and from the $\kappa$-uniform approximation [Eq.~(\ref{eq:E_tau_only})] separately, we quantify the relative contributions of $\kappa$ and $\tau$ variations to the total integrability error.

For the 50-chain PDB dataset, we computed $E[n]$ at every interior residue and characterized its statistical properties: the chain-averaged error $\langle E\rangle$, the spatial concentration (Gini coefficient of the $E[n]$ profile), and the correlation of $\langle E\rangle$ with chain length $N$, helical fraction, and the standard deviations $\sigma_\kappa$ and $\sigma_\tau$ of the geometric variables.

\subsection{Dispersion-relation prediction}
\label{sec:prediction}

The uniform-segment dispersion relation derived in Ref.~\cite{wang2026Structural}
\begin{equation}
	\cos\tau = 1 + \frac{V_{\text{re}}}{2\beta}
	\label{eq:dispersion}
\end{equation}
provides a one-to-one mapping between $V_{\text{re}}$ and $|\tau|$ within any segment where $\kappa$ and $\tau$ are approximately constant. We exploit this relation to predict helical backbone structures from the mean geometric parameters alone, without any sequence information or energy function.

We evaluate the baseline validity of the framework through a full-chain prediction protocol in which we compute the mean curvature $\langle\kappa\rangle$ and mean torsion $\langle\tau\rangle$ from the experimental structure. A uniform helix with these parameters is constructed by assigning $\kappa[n] = \langle\kappa\rangle$ and $\tau[n] = \langle\tau\rangle$ at every residue and reconstructing the C$_\alpha$ backbone via the discrete Frenet equations. The initial Frenet frame defined by the position $\mathbf{r}[1]$, tangent $\hat{\mathbf{t}}[1]$, and normal $\hat{\mathbf{n}}[2]$ is extracted from the first three C$_\alpha$ atoms of the experimental structure while the predicted and experimental structures are compared by the root-mean-square deviation (RMSD) after optimal superposition via the Kabsch algorithm~\cite{kabsch1976solution}.

We address local geometric heterogeneity through a segmented prediction strategy where the integrability error profile $E[n]$ serves to identify residues where the integrable approximation breaks down. We designate residues with $E[n] > E_{\text{cut}}$ as cut points to partition the chain into segments within which $E[n] \leq E_{\text{cut}}$ and predict each segment independently using its own $\langle\kappa\rangle_{\text{seg}}$ and $\langle\tau\rangle_{\text{seg}}$ with the initial Frenet frame extracted from the corresponding experimental segment. The segment-level predictions are concatenated and the overall RMSD is computed against the full experimental structure after Kabsch alignment. We tested thresholds $E_{\text{cut}} \in \{0.03, 0.05, 0.10, 0.15, 0.20, 0.30\}$ and identified $E_{\text{cut}} = 0.10$ as the optimal trade-off between prediction accuracy and chain coverage as detailed in Sec.~\ref{sec:results_prediction}.

To characterize the applicability boundary of the prediction, we classified chains into four zones based on the mean error and torsion standard deviation. Zone~A is defined by the parameter range $\langle E\rangle < 0.10$ and $\sigma_\tau < 0.40$\,rad whereas Zone~B captures chains where $\langle E\rangle < 0.10$ but $\sigma_\tau \geq 0.40$\,rad. Zone~C ($0.10 \leq \langle E\rangle < 0.20$) contains 7 chains with 71\% success and median RMSD $= 0.54$\,\AA. Zone~D ($\langle E\rangle \geq 0.20$) contains 2 chains with 0\% success (median RMSD $= 3.94$\,\AA) and defines a limit where the density of integrability violations precludes effective reconstruction via segmentation. This classification establishes a quantitative criterion to determine \textit{a priori} whether a given helical peptide falls within the integrable regime.

\section{Results}
\label{sec:results}

\subsection{Information-theoretic properties of the \texorpdfstring{$(\varphi,\psi)\to(\kappa,\tau)$}{(phi,psi)->(kappa,tau)} mapping}
\label{sec:results_mapping}

\begin{figure*}[t]
\includegraphics[width=0.9\textwidth]{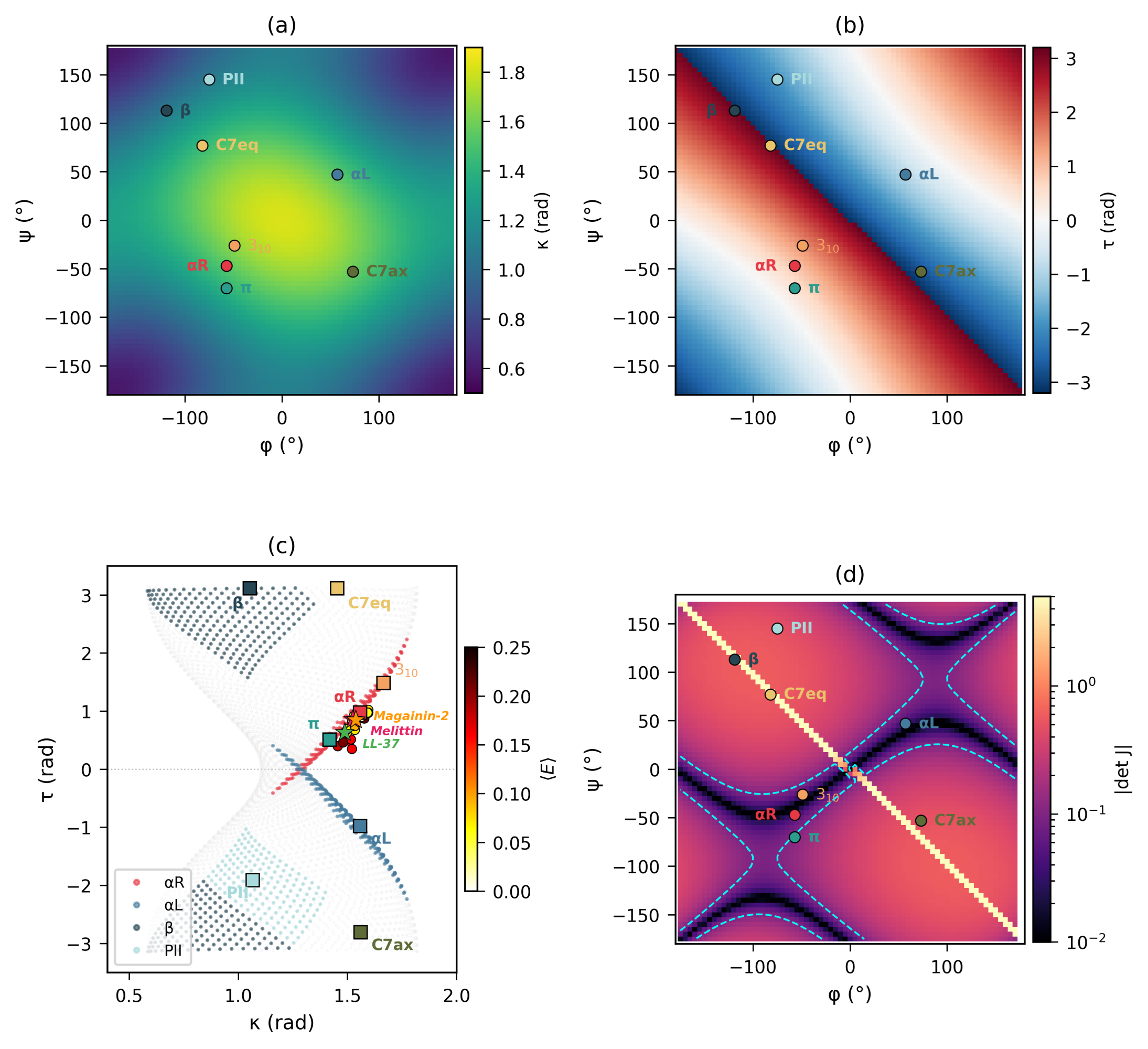}
\caption{The $(\varphi,\psi)\to(\kappa,\tau)$ mapping.
	(a)~Curvature $\kappa(\varphi,\psi)$ across the Ramachandran plane, with
	eight standard conformations marked. The minimum
	$\kappa_{\min}=0.58$\,rad is far from zero, ruling out the $\kappa\to 0$
	coordinate singularity.
	(b)~Torsion $\tau(\varphi,\psi)$, showing the antisymmetric chirality
	relation $\tau(\varphi,\psi)=-\tau(-\varphi,-\psi)$ and the sign-change
	boundary along the diagonal.
	(c)~The Hasimoto--Ramachandran space: four Ramachandran regions
	($\alpha_R$, $\alpha_L$, $\beta$, PII) mapped into $(\kappa,\tau)$
	coordinates, with the 50 PDB helical chains overlaid as circles colored
	by $\langle E\rangle$. Three core antimicrobial peptides are highlighted
	with colored stars. The $\beta$ and PII regions share 89.9\% of their
	$\kappa$ range but are offset in $\tau$ (median nearest-neighbor distance
	0.29\,rad).
	(d)~Jacobian determinant $|\det J|$ on a logarithmic scale with the
	$|J|=0.1$ contour in cyan dashed, revealing the ill-conditioned
	$\alpha_R$ basin (median condition number $\sim 31$, 68.4\% of points
	below $|J|=0.1$).}
\label{fig:mapping}
\end{figure*}

The $(\varphi,\psi)\to(\kappa,\tau)$ mapping, constructed by scanning the Ramachandran plane at $5^\circ$ resolution through the NeRF pipeline (Sec.~\ref{sec:mapping}), reveals a strongly nonlinear but globally information-preserving coordinate transformation. Figure~\ref{fig:mapping} summarizes the key aspects of this mapping and Table~\ref{tab:mapping} collects the quantitative results. Under standard peptide-bond geometry the curvature spans $\kappa\in[0.58,\,1.82]$\,rad across the entire Ramachandran plane [Fig.~\ref{fig:mapping}(a)], with the minimum $\kappa_{\min}=0.58$\,rad occurring at the fully extended conformation $(\varphi,\psi)\approx(-180^\circ,180^\circ)$. Because this lower bound is far from zero, the coordinate singularity at $\kappa=0$ where the Frenet frame becomes undefined and the Hasimoto phase $\arg\psi$ is discontinuous is geometrically inaccessible for any backbone conformation built from standard peptide bonds. This observation rules out $\kappa\to 0$ as the microscopic origin of the $2^N$ chirality degeneracy identified in Paper~I. At the global level the mapping preserves the vast majority of conformational information. The Shannon entropy of the Ramachandran distribution is $H(\varphi,\psi)=11.12$\,bits while the entropy of the image distribution in $(\kappa,\tau)$ space is $H(\kappa,\tau)=10.38$\,bits, yielding an information retention ratio of $H(\kappa,\tau)/H(\varphi,\psi)=93.4\%$.

Despite this high global information retention the mapping exhibits severe local ill-conditioning in the $\alpha$-helical region. The Jacobian determinant $|J|=|\partial(\kappa,\tau)/\partial(\varphi,\psi)|$ drops below 0.1 over 68.4\% of the $\alpha_R$ basin [Fig.~\ref{fig:mapping}(d)] with a median condition number of 31 which implies that large changes in $(\varphi,\psi)$ within the helical basin yield only small changes in $(\kappa,\tau)$ and that the mapping compresses the helical region of Ramachandran space into a narrow strip in Frenet space. The $\alpha_L$ basin exhibits identical statistics by the chirality symmetry $\kappa(\varphi,\psi)=\kappa(-\varphi,-\psi)$, $\tau(\varphi,\psi)=-\tau(-\varphi,-\psi)$, whereas the $\beta$ and polyproline-II (PII) regions have well-conditioned Jacobians (median condition numbers 3.6 and 3.9 respectively) with no points below $|J|=0.1$. This Jacobian condition constitutes the microscopic mechanism underlying the chirality degeneracy inherent to the discrete mapping~\cite{wang2026Structural} where the many-to-one character of the transformation in the helical region causes multiple distinct backbone conformations to project onto nearly identical Frenet parameters. A further consequence of the nonlinear compression is that the $\beta$ and polyproline-II (PII) regions well separated in $(\varphi,\psi)$ space share 89.9\% of their $\kappa$ range when projected into $(\kappa,\tau)$ space [Fig.~\ref{fig:mapping}(c)]. Although the two regions are offset in $\tau$ by a median nearest-neighbor distance of 0.29\,rad, their curvature degeneracy means that $\kappa$ alone cannot distinguish them, explaining the insensitivity of $V_{\text{re}}$ to sequence identity~\cite{wang2026Structural} where $\beta$ and PII conformations produce nearly identical curvature profiles such that any sequence-dependent preference between them is largely invisible to the Hasimoto representation.

Within this landscape the 50 PDB helical chains cluster tightly near the $\alpha_R$ centroid ($\langle\kappa\rangle=1.50\pm0.14$\,rad, $\langle\tau\rangle=0.83\pm0.57$\,rad) as shown in Fig.~\ref{fig:mapping}(c). This clustering places the dataset in a region where $\kappa$ is large and approximately uniform, so that the integrability conditions are most nearly satisfied. The question then becomes whether the spatial non-uniformity of the Frenet parameters along each chain is small enough for the analytic dispersion relation to remain predictive, a question addressed by the integrability error analysis that follows.

\begin{table}[t]
\caption{Properties of the $(\varphi,\psi)\to(\kappa,\tau)$ mapping under standard peptide-bond geometry. The Ramachandran plane was scanned at $5^\circ$ resolution ($72\times72$ grid). Region statistics use the standard basin definitions: $\alpha_R$ ($\varphi\in[-100^\circ,-30^\circ]$, $\psi\in[-80^\circ,-10^\circ]$), $\alpha_L$ (mirror image), $\beta$ ($\varphi\in[-180^\circ,-100^\circ]$, $\psi\in[80^\circ,180^\circ]$), PII ($\varphi\in[-100^\circ,-50^\circ]$, $\psi\in[100^\circ,180^\circ]$).}
\label{tab:mapping}
\begin{ruledtabular}
\begin{tabular}{lc}
\textrm{Quantity} & \textrm{Value} \\
\colrule
$\kappa$ range (full Ramachandran) & $[0.58,\,1.82]$\,rad \\
$\kappa_{\min}$ & 0.58\,rad \\
$H(\varphi,\psi)$ & 11.12\,bits \\
$H(\kappa,\tau)$ & 10.38\,bits \\
Information retention & 93.4\% \\
\colrule
\multicolumn{2}{c}{\textit{Region-specific Jacobian properties}} \\
\colrule
$\alpha_R$: median cond($J$) & 31 \\
$\alpha_R$: fraction $|J|<0.1$ & 68.4\% \\
$\beta$: median cond($J$) & 3.6 \\
PII: median cond($J$) & 3.9 \\
$\beta$--PII $\kappa$-range overlap & 89.9\% \\
$\beta$--PII nearest-neighbor median & 0.29\,rad \\
\end{tabular}
\end{ruledtabular}
\end{table}

\subsection{Analytic theory and PDB validation of the integrability error}
\label{sec:results_integrability}

Equations~(\ref{eq:E_simplified})--(\ref{eq:E_tau_only}) establish that $E[n]$ vanishes identically whenever $\kappa$ and $\tau$ are spatially constant, so that a perfectly uniform helix satisfies the discrete dispersion relation at every site. Integrability breaking therefore originates entirely from spatial non-uniformity of the geometric variables rather than from any intrinsic deficiency of the Hasimoto formalism. Across the 50-chain PDB dataset the chain-averaged integrability error spans $\langle E\rangle \in [0.014,\,0.216]$ with a median of 0.050 (Table~\ref{tab:integrability}), and $\langle E\rangle$ shows no significant correlation with chain length ($r = 0.13$, $p = 0.38$). Integrability is thus a local property determined by the residue-level geometric environment rather than a global property that accumulates with chain size. This length-independence, together with the information-preserving character of the mapping established in Sec.~\ref{sec:results_mapping}, provides the theoretical prerequisite for the segmented prediction strategy developed below.

\begin{figure*}[t]
\includegraphics[width=0.9\textwidth]{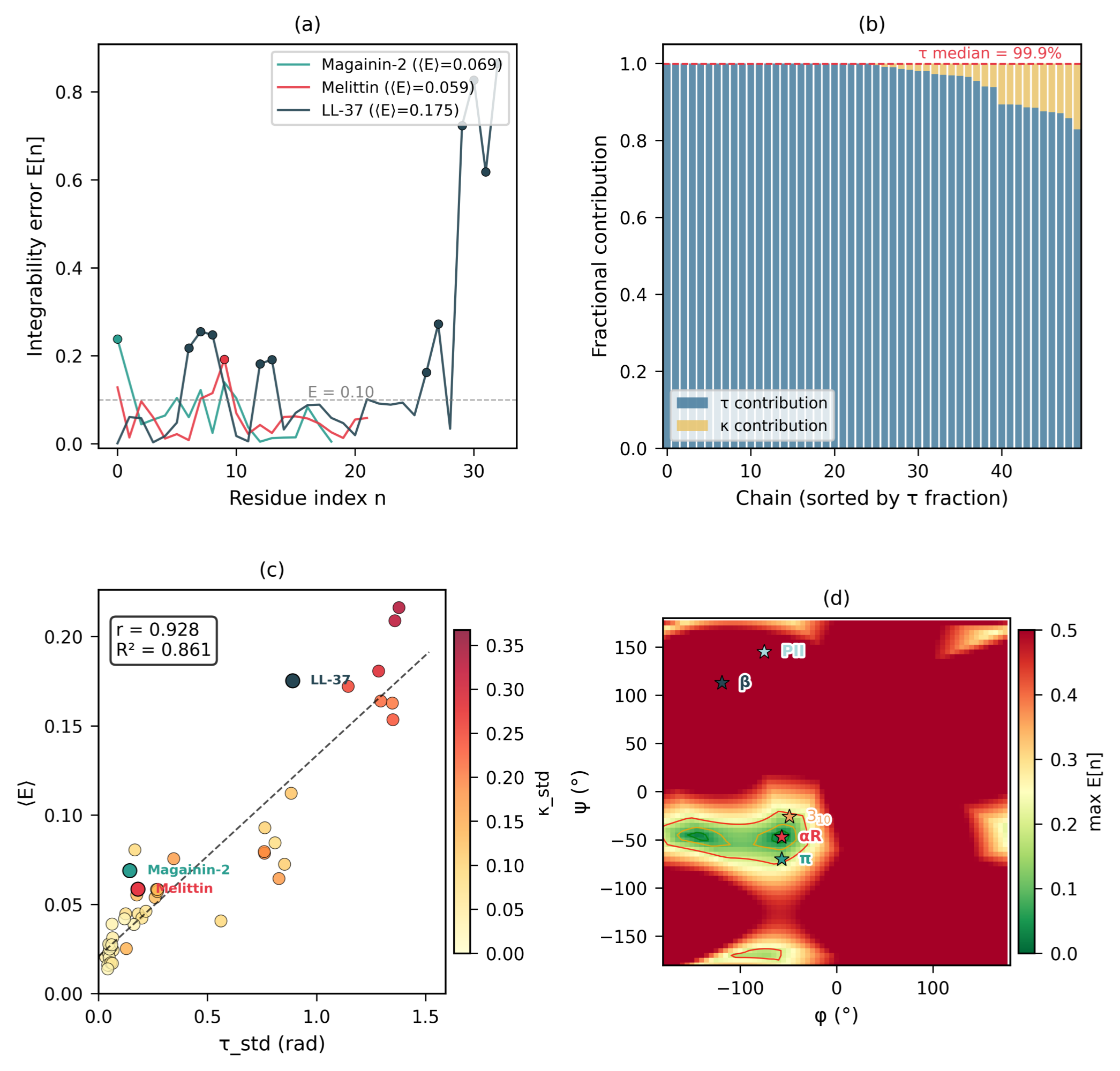}
\caption{Integrability error analysis.
	(a)~$E[n]$ profiles for the three core antimicrobial peptides:
	Magainin-2 ($\langle E\rangle = 0.069$), Melittin
	($\langle E\rangle = 0.059$), and LL-37
	($\langle E\rangle = 0.175$). The dashed line marks the segmentation
	threshold $E = 0.10$; filled circles highlight residues with $E > 0.15$.
	LL-37 exhibits a dominant spike at indices 29--32 ($E > 0.5$)
	corresponding to a $\tau$ sign reversal at the C-terminal kink, while
	Magainin-2 and Melittin maintain low-amplitude profiles consistent with
	high helical regularity.
	(b)~Decomposition of $\langle E\rangle$ into $\tau$ (blue) and $\kappa$
	(yellow) contributions for all 50 chains, sorted by $\tau$ fraction. The
	dashed line marks the median $\tau$ contribution of 99.9\%.
	(c)~$\langle E\rangle$ versus $\sigma_\tau$ for the 50 chains (colored
	by $\sigma_\kappa$), with linear regression ($r = 0.928$,
	$R^2 = 0.861$), indicating that torsion non-uniformity alone explains
	over 86\% of the variance in integrability error.
	(d)~Integrability tolerance map: maximum $E[n]$ resulting from mutating
	the central residue of a 10-residue $\alpha_R$ helix to every
	$(\varphi, \psi)$ point on a $5^\circ$ grid. Contours at $E = 0.05$
	(green), 0.10 (orange), and 0.20 (red) delineate the integrable region;
	standard conformations are marked by colored stars. The $E < 0.10$
	region occupies only 1.2\% of the Ramachandran plane, confined to a
	narrow neighborhood of the $\alpha_R$ basin.}
\label{fig:integrability}
\end{figure*}

A systematic decomposition reveals that torsion non-uniformity is the dominant source of integrability breaking. A single-variable linear regression of $\langle E\rangle$ on $\sigma_\tau$ yields $R^2 = 0.861$ [Fig.~\ref{fig:integrability}(c)], and adding $\sigma_\kappa$ as a second predictor improves the fit only marginally ($R^2 = 0.897$, $\Delta R^2 = 0.036$). The $\kappa$-uniform decomposition [Eq.~(\ref{eq:E_tau_only})] provides a more direct test. Computing $E[n]$ with $\kappa$ held at its chain mean while retaining the actual $\tau[n]$ profile reproduces a median 99.9\% of the full integrability error [Fig.~\ref{fig:integrability}(b)], so that the independent contribution of $\kappa$ variation is negligible for all practical purposes. This $\tau$-dominance has a clear physical origin. In the $\beta^{\pm}$-constant regime $V_{\text{re}}$ depends on $\tau$ through $\cos\tau$ [Eq.~(\ref{eq:Vre})], and the curvature ratios $r^{\pm} = \kappa[n{\pm}1]/\kappa[n]$ remain close to unity in helical regions ($\sigma_\kappa/\langle\kappa\rangle \approx 0.07$ for the dataset median). The curvature is structurally rigid while the torsion carries essentially all of the conformational flexibility that drives deviations from integrability.

The spatial distribution of $E[n]$ along each chain is far from uniform. The $E[n]$ profiles of the three core systems [Fig.~\ref{fig:integrability}(a)] show that integrability breaking concentrates at a few localized sites while the majority of residues remain near-integrable. The median Gini coefficient of the $E[n]$ distribution across the 50 chains is 0.55, quantifying this spatial inequality. For LL-37 ($\langle E\rangle = 0.175$) the profile shows a moderate cluster near the mid-chain kink (profile indices $n=6$--9 and 12--13, $E\sim 0.13$--$0.26$) and a dominant spike at the C-terminal boundary (indices $n=29$--32, $E > 0.5$) where $\tau$ undergoes a sign reversal, while the intervening helical segments maintain $E[n] < 0.05$. Magainin-2 ($\langle E\rangle = 0.069$) and melittin ($\langle E\rangle = 0.059$) exhibit more uniform low-amplitude profiles consistent with their higher helical regularity. This spatial concentration provides the physical basis for isolating integrable islands by cutting the chain at high-$E$ sites. The sensitivity of integrability to conformational heterogeneity is further quantified by a tolerance map [Fig.~\ref{fig:integrability}(d)] constructed by mutating the central residue of a 10-residue $\alpha_R$ helix ($\varphi = -57^\circ$, $\psi = -47^\circ$) to every $(\varphi,\psi)$ point on a $5^\circ$ grid. The region satisfying $E < 0.10$ occupies only ${\sim}1.2\%$ of the Ramachandran plane, confined to a narrow neighborhood of the $\alpha_R$ basin. Even the $3_{10}$ and $\pi$-helix conformations, which are structurally similar to $\alpha_R$, produce $E > 0.10$ when inserted as single-residue perturbations. This extreme sensitivity explains why integrability is restricted to conformationally homogeneous helical segments and why non-helical insertions act as natural boundaries of integrable islands, motivating the segmented prediction strategy presented next.

\begin{table}[b]
	\caption{Integrability error statistics for the 50-chain helical peptide dataset. $\sigma_\tau$ and $\sigma_\kappa$ denote the intra-chain standard deviations of torsion and curvature, respectively.}
	\label{tab:integrability}
	\begin{ruledtabular}
		\begin{tabular}{ll}
			\multicolumn{2}{c}{\textit{Global statistics}} \\
			\colrule
			$\langle E\rangle$ range & $[0.014,\,0.216]$ \\
			$\langle E\rangle$ median & 0.050 \\
			$\langle E\rangle$ vs chain length $N$ & $r = 0.13$ ($p = 0.38$) \\
			\colrule
			\multicolumn{2}{c}{\textit{Regression analysis}} \\
			\colrule
			$\sigma_\tau$ single-variable $R^2$ & 0.861 \\
			$(\sigma_\kappa + \sigma_\tau)$ two-variable $R^2$ & 0.897 \\
			$\sigma_\kappa$ independent contribution $\Delta R^2$ & 0.036 \\
			\colrule
			\multicolumn{2}{c}{\textit{$\tau$ dominance ($\kappa$-uniform decomposition)}} \\
			\colrule
			$\tau$ contribution (median) & 99.9\% \\
			Gini coefficient of $E[n]$ (median) & 0.55 \\
			\colrule
			\multicolumn{2}{c}{\textit{Integrability tolerance}} \\
			\colrule
			$E < 0.10$ area (Ramachandran) & 1.2\% \\
			\colrule
			\multicolumn{2}{c}{\textit{Core systems}} \\
			\colrule
			Melittin (2MLT, 26 AA) & $\langle E\rangle = 0.059$, $\sigma_\tau = 0.18$ \\
			Magainin-2 (2MAG, 23 AA) & $\langle E\rangle = 0.069$, $\sigma_\tau = 0.14$ \\
			LL-37 (2K6O, 37 AA) & $\langle E\rangle = 0.175$, $\sigma_\tau = 0.89$ \\
		\end{tabular}
	\end{ruledtabular}
\end{table}

\subsection{Dispersion-relation prediction and applicability boundaries}
\label{sec:results_prediction}

\begin{figure*}[tb]
\includegraphics[width=0.9\textwidth]{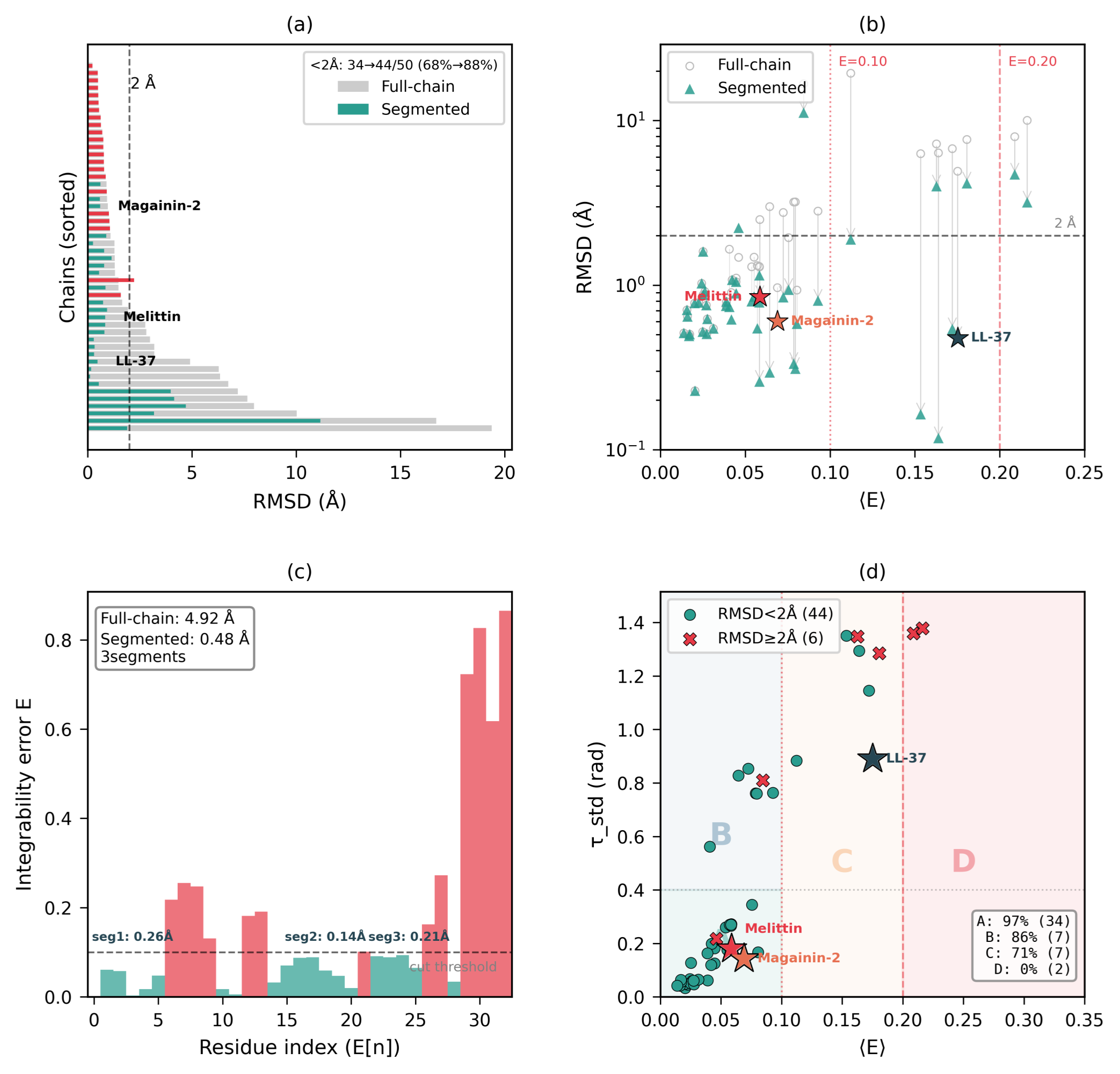}
\caption{Dispersion-relation prediction and applicability boundaries.
	(a)~Paired RMSD comparison for all 50 chains, sorted by full-chain RMSD.
	Gray bars show full-chain prediction from chain-averaged
	$(\langle\kappa\rangle, \langle\tau\rangle)$; colored bars show
	segmented prediction ($E > 0.10$ cutoff, evaluated exclusively over retained integrable islands $\ge 4$ residues), with green indicating
	improvement and red indicating worsening. The dashed line marks the
	2\,\AA\ success threshold. Segmentation raises the success rate from
	68\% (34/50) to 88\% (44/50).
	(b)~RMSD versus $\langle E\rangle$ on a logarithmic scale. Open circles
	represent full-chain predictions and filled triangles represent segmented
	predictions. Arrows connect pairs with large improvement and lines
	connect pairs with moderate improvement; pairs with negligible change are
	left unconnected. Vertical dashed lines mark the zone boundaries at
	$\langle E\rangle = 0.10$ and 0.20, and the horizontal dashed line marks
	the 2\,\AA\ threshold. Colored stars highlight the three core systems:
	Melittin ($2.50 \to 0.85$\,\AA), Magainin-2 ($0.96 \to 0.60$\,\AA),
	and LL-37 ($4.92 \to 0.48$\,\AA).
	(c)~LL-37 case study: residue-level $E[n]$ profile, with red bars
	indicating high-$E$ sites ($E > 0.10$) that serve as segmentation cut
	points. The dashed line marks the cut threshold $E = 0.10$.
	Segment-level RMSD values are annotated, demonstrating that each
	integrable segment achieves sub-angstrom accuracy (seg1: 0.26\,\AA,
	seg2: 0.14\,\AA, seg3: 0.21\,\AA) despite the full-chain RMSD of
	4.92\,\AA.
	(d)~Four-zone applicability map in the $(\langle E\rangle, \sigma_\tau)$
	plane. Circles denote chains with segmented RMSD $< 2$\,\AA\ (44
	chains) and crosses denote failures (6 chains). Background shading
	indicates Zone~A ($\langle E\rangle < 0.10$, $\sigma_\tau < 0.40$; 34
	chains, 97\% success), Zone~B ($\langle E\rangle < 0.10$,
	$\sigma_\tau \geq 0.40$; 7 chains, 86\%), Zone~C
	($0.10 \leq \langle E\rangle < 0.20$; 7 chains, 71\%), and Zone~D
	($\langle E\rangle \geq 0.20$; 2 chains, 0\%). Zone~B (median
	0.74\,\AA) achieves comparable sub-angstrom accuracy on its integrable cores to the globally uniform Zone~A (0.77\,\AA), supporting the segment-level
	integrable-island interpretation.}
\label{fig:prediction}
\end{figure*}

The integrability error analysis of Sec.~\ref{sec:results_integrability} established that deviations from the integrable limit are local, spatially concentrated, and dominated by torsion non-uniformity. These properties suggest that the analytic dispersion relation (Sec.~\ref{sec:prediction}) should predict helical backbone coordinates with high accuracy within segments of low $E[n]$, even when the chain as a whole deviates from uniformity. We test this hypothesis on all 50 chains using two strategies and delineate the quantitative applicability boundary. Figure~\ref{fig:prediction} displays the results and Table~\ref{tab:prediction} collects the statistics.

As a baseline we first apply full-chain prediction, reconstructing a uniform helix from the chain-averaged parameters $(\langle\kappa\rangle, \langle\tau\rangle)$ via the dispersion relation $\cos\tau = 1 + V_{\text{re}}/(2\beta)$ and computing the C$_\alpha$ RMSD after Kabsch alignment. Of the 50 chains 34 (68\%) achieve RMSD $< 2$\,\AA\ with a median of 1.29\,\AA\ [Fig.~\ref{fig:prediction}(a), gray bars]. The best single predictor of full-chain RMSD is $\sigma_\tau$ ($r = 0.732$), followed by $\langle E\rangle$ ($r = 0.662$), confirming the $\tau$-dominance identified in Sec.~\ref{sec:results_integrability}. Chains with low torsion heterogeneity ($\sigma_\tau < 0.20$\,rad) are well approximated by a single uniform helix, whereas chains with large $\sigma_\tau$ deviate substantially because the uniform-helix assumption breaks down at residues where $\tau$ fluctuates.

The $E[n]$ profile, established above as a residue-level integrability map, provides a natural segmentation criterion that substantially improves upon the full-chain baseline. Cutting the chain at every site where $E[n] > 0.10$ partitions the backbone into contiguous fragments. Crucially, segments shorter than four residues are discarded, as these typically correspond to non-integrable structural defects or terminal fraying. By evaluating the RMSD exclusively over these retained integrable islands and predicting each independently using its own $(\langle\kappa\rangle_{\text{seg}}, \langle\tau\rangle_{\text{seg}})$, the success rate rises from 68\% to 88\% (44/50 chains with island RMSD $< 2$\,\AA), and the median RMSD is reduced from 1.29\,\AA\ to 0.77\,\AA\ [Fig.~\ref{fig:prediction}(a), colored bars]. Of the 50 chains 28 improve under segmentation and only one worsens (6YTU\_A, discussed below). The improvement is most dramatic for chains that fail under full-chain prediction, where the median RMSD of the 16 initially failing chains drops from 6.32\,\AA\ to 0.82\,\AA. LL-37 (2K6O, 37 residues) exemplifies this gain [Fig.~\ref{fig:prediction}(c)]. Its $E[n]$ profile reveals a cluster of four consecutive high-$E$ residues ($E > 0.5$) near positions 29--32 coinciding with a kink region where $\tau$ undergoes a sign reversal. Full-chain prediction yields RMSD $= 4.92$\,\AA\ which represents a substantial deviation driven by this localized non-integrability. Segmentation at $E > 0.10$ isolates the kink and the resulting integrable segments are each predicted to sub-angstrom accuracy yielding a composite RMSD of 0.48\,\AA\ which corresponds to a 90\% improvement. The two other core systems also benefit, with melittin improving from 2.50\,\AA\ to 0.85\,\AA\ and magainin-2 from 0.96\,\AA\ to 0.60\,\AA\ (Table~\ref{tab:prediction}).

Plotting segmented RMSD against $\langle E\rangle$ and $\sigma_\tau$ reveals a natural four-zone classification of the applicability boundary [Fig.~\ref{fig:prediction}(d)]. Zone~A ($\langle E\rangle < 0.10$, $\sigma_\tau < 0.40$\,rad) contains 34 chains with a 97\% success rate and a median RMSD of $0.77$\,\AA, representing the core integrable regime where even full-chain prediction largely succeeds. Zone~B ($\langle E\rangle < 0.10$, $\sigma_\tau \geq 0.40$\,rad) contains 7 chains with an 86\% success rate and an exceptional median RMSD of 0.74\,\AA. Rather than being counterintuitive, this highlights the physical precision of the $E[n]$ metric. Chains in Zone~B typically suffer from severe terminal fraying or isolated kinks that ruin the full-chain prediction (median full-chain RMSD $= 3.00$\,\AA). The segmentation algorithm acts as a stringent filter: it selectively trims these non-integrable boundary residues. Once this noise is discarded, the remaining core often consists of a single, highly pure integrable island that achieves sub-angstrom accuracy, rivaling the globally uniform chains of Zone~A. Zone~C ($0.10 \leq \langle E\rangle < 0.20$) contains 7 chains with 71\% success and median RMSD $= 0.54$\,\AA. Zone~D ($\langle E\rangle \geq 0.20$) contains 2 chains with 0\% success (median RMSD $= 3.94$\,\AA) and represents a practical limit where the density of integrability violations precludes effective reconstruction via segmentation. The counterintuitive superiority of Zone~B over Zone~A provides evidence for the central concept of this work as full-chain prediction shows significant deviation for 6 of 7 Zone~B chains (median RMSD $= 3.00$\,\AA) yet segmentation resolves 6 of 7 to below 2\,\AA\ (median $= 0.74$\,\AA) which demonstrates that the integrable islands within these chains are among the most uniform segments in the entire dataset. A chain can thus harbor both highly integrable islands and distinct non-integrable barriers which the $E[n]$ profile distinguishes.

One chain (6YTU\_A) worsens under segmentation because an unnecessary cut produces a 3-residue terminal fragment too short for reliable prediction, with its full-chain RMSD of 1.47\,\AA\ increasing to 2.23\,\AA. This motivates a two-step strategy in which full-chain prediction is attempted first and segmented prediction is applied only if the RMSD exceeds 2\,\AA. Under this protocol no chain in the dataset is made worse by segmentation and the overall success rate reaches 90\% (45/50). The optimal segmentation threshold is $E > 0.10$, which maximizes the $F_1$ score ($F_1 = 0.881$) in the precision-coverage tradeoff. Lower thresholds such as $E > 0.03$ achieve 100\% accuracy on retained segments but discard 46\% of residues, while higher thresholds such as $E > 0.20$ retain more residues but miss critical non-integrable sites. With the applicability boundary now quantitatively defined, we turn to the converse question of whether the framework can guide the inverse design of helical backbones within the integrable regime.

\begin{table}[b]
	\caption{Dispersion-relation prediction results for the 50-chain helical peptide dataset. Full-chain: prediction from chain-averaged $(\langle\kappa\rangle, \langle\tau\rangle)$. Segmented: $E > 0.10$ cutoff with independent segment prediction. Zones are defined by $\langle E\rangle$ and $\sigma_\tau$ thresholds (see text).}
	\label{tab:prediction}
	\begin{ruledtabular}
		\begin{tabular}{lcc}
			\multicolumn{3}{c}{\textit{Overall performance}} \\
			\colrule
			& Full-chain & Segmented \\
			\colrule
			Median RMSD (\AA) & 1.29 & 0.77 \\
			RMSD $< 2$\,\AA\ success rate & 68\% (34/50) & 88\% (44/50) \\
			\colrule
			\multicolumn{3}{c}{\textit{Four-zone analysis (segmented prediction)}} \\
			\colrule
			Zone & Success rate & Median RMSD (\AA) \\
			\colrule
			A: $\langle E\rangle < 0.10$, $\sigma_\tau < 0.40$ ($n=34$) & 97\% & 0.77 \\
			B: $\langle E\rangle < 0.10$, $\sigma_\tau \geq 0.40$ ($n=7$) & 86\% & 0.74 \\
			C: $0.10 \leq \langle E\rangle < 0.20$ ($n=7$) & 71\% & 0.54 \\
			D: $\langle E\rangle \geq 0.20$ ($n=2$) & 0\% & 3.94 \\
			\colrule
			\multicolumn{3}{c}{\textit{Core systems (full-chain $\to$ segmented)}} \\
			\colrule
			Melittin (2MLT, 26 AA) & \multicolumn{2}{c}{$2.50 \to 0.85$\,\AA} \\
			Magainin-2 (2MAG, 23 AA) & \multicolumn{2}{c}{$0.96 \to 0.60$\,\AA} \\
			LL-37 (2K6O, 37 AA) & \multicolumn{2}{c}{$4.92 \to 0.48$\,\AA} \\
			\colrule
			\multicolumn{3}{c}{\textit{Predictors of full-chain RMSD}} \\
			\colrule
			$r(\langle E\rangle, \text{RMSD})$ & \multicolumn{2}{c}{0.662} \\
			$r(\sigma_\tau, \text{RMSD})$ & \multicolumn{2}{c}{0.732} \\
		\end{tabular}
	\end{ruledtabular}
\end{table}

\subsection{Inverse design feasibility}
\label{sec:results_inverse}

\begin{figure*}[tb]
\includegraphics[width=0.9\textwidth]{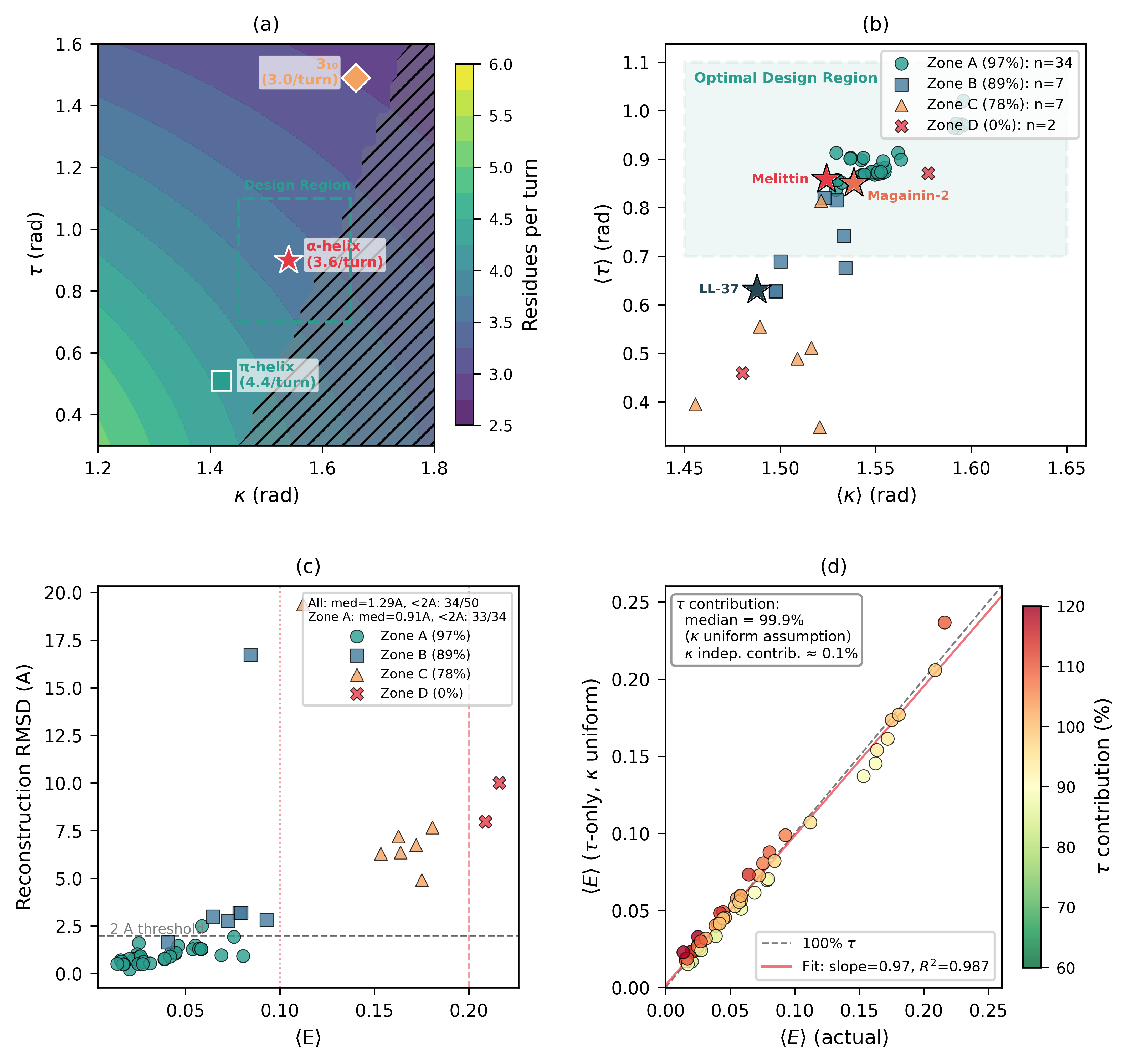}
\caption{Inverse design feasibility. (a)~Helical parameter space $(\kappa, \tau)$ with $n_{\text{per turn}}$ contours (gray). Blue shading marks the Ramachandran-reachable region (68.2\% of the scanned grid); stars indicate canonical helix types ($\alpha$, $3_{10}$, $\pi$). (b)~The 50 PDB chains in $(\langle\kappa\rangle, \langle\tau\rangle)$ space colored by Zone classification. The dashed rectangle marks the suggested optimal design region defined by $\kappa \in [1.45, 1.65]$ and $\tau \in [0.70, 1.10]$. (c)~Uniform-helix reconstruction RMSD versus $\langle E\rangle$ for all 50 chains; Zone~A chains (filled circles) cluster at low RMSD. (d)~$\tau$ dominance: $E_{\tau\text{-only}}$ (computed under the $\kappa$-uniform assumption) versus $E_{\text{actual}}$ for all 50 chains. The near-unity slope (0.969) and $R^2 = 0.987$ confirm that torsion non-uniformity accounts for $\sim$99\% of integrability breaking.}
\label{fig:inverse}
\end{figure*}

The applicability zones defined in Sec.~\ref{sec:results_prediction} delineate where the dispersion relation predicts helical structures from mean Frenet parameters. The converse question follows naturally. Given a target helix geometry $(\kappa_0, \tau_0)$, can the Hasimoto framework guide the design of a peptide backbone that realizes it? Figure~\ref{fig:inverse} and Table~\ref{tab:inverse} summarize the analysis. The starting point is the exact self-consistency of the reverse dispersion relation. For a perfectly uniform helix with constant $(\kappa_0, \tau_0)$ the integrability error vanishes identically, $E[n] \equiv 0$ for all $n$, because $\kappa[n] = \kappa_0$ and $\tau[n] = \tau_0$ render the effective potential spatially constant. This is not an approximation but an exact algebraic identity following from Eqs.~(\ref{eq:Vre})--(\ref{eq:Vim}). The predicted effective potential $V_{\text{re}} = 2\beta(\cos\tau_0 - 1)$ matches the actual value to machine precision ($< 10^{-15}$) for all tested configurations including the $\alpha$-helix ($\kappa_0 = 1.54$, $\tau_0 = 0.90$), the $3_{10}$-helix ($\kappa_0 = 1.66$, $\tau_0 = 1.49$), and the $\pi$-helix ($\kappa_0 = 1.42$, $\tau_0 = 0.51$), with reconstructed helices reproducing the input coordinates at RMSD $< 10^{-14}$\,\AA. However not all $(\kappa_0, \tau_0)$ pairs correspond to physically realizable peptide backbones. Scanning a $50 \times 50$ grid over $\kappa \in [1.2, 1.8]$\,rad and $\tau \in [0.3, 1.6]$\,rad and checking whether each target maps back to a $(\varphi, \psi)$ pair within the allowed Ramachandran region shows that 1706 of 2500 grid points (68.2\%) are Ramachandran-reachable [Fig.~\ref{fig:inverse}(a)]. The reachable region spans $\kappa \in [1.20, 1.75]$\,rad and $\tau \in [0.30, 1.60]$\,rad, covering helices from $n_{\text{per turn}} \approx 2.8$ ($3_{10}$-like) to $\approx 5.1$ ($\pi$-like), and all three canonical helix types fall within it with the $\alpha$-helix at $n_{\text{per turn}} = 3.62$ (experimental 3.6), the $3_{10}$-helix at 2.98 (experimental 3.0), and the $\pi$-helix at 4.20 (experimental 4.4). The 50 PDB chains cluster tightly near the $\alpha$-helix centroid ($\langle\kappa\rangle = 1.53 \pm 0.03$\,rad, $\langle\tau\rangle = 0.78 \pm 0.17$\,rad) with Zone~A chains occupying a narrowly defined optimal design interval at $\kappa \in [1.52, 1.60]$ and $\tau \in [0.84, 1.02]$ as illustrated in Fig.~\ref{fig:inverse}(b).

To quantify the gap between the idealized design target and real peptide structures we reconstructed a uniform helix for each of the 50 chains using its mean parameters $(\langle\kappa\rangle, \langle\tau\rangle)$ and the initial Frenet frame from the experimental structure. The median reconstruction RMSD is 1.29\,\AA\ across all 50 chains with 34/50 (68\%) achieving RMSD $< 2$\,\AA\ [Fig.~\ref{fig:inverse}(c)]. Within Zone~A ($\langle E\rangle < 0.10$, $\sigma_\tau < 0.40$) the median improves to 0.91\,\AA\ with 33/34 (97\%) below 2\,\AA. The reconstruction RMSD correlates strongly with $\sigma_\tau$ ($r = 0.732$) and $\langle E\rangle$ ($r = 0.662$), confirming that the dominant source of deviation from the design target is torsion non-uniformity in the real structure rather than curvature variation. This result connects directly to the $\tau$-dominance established in Sec.~\ref{sec:results_integrability} and reinforces the conclusion that the integrability boundary identified in Sec.~\ref{sec:results_prediction} simultaneously defines the feasibility boundary for inverse design.

The decomposition of $E[n]$ into $\kappa$ and $\tau$ contributions translates into a concrete design principle. Under the $\kappa$-uniform assumption ($\kappa[n] = \langle\kappa\rangle$ for all $n$) the $\tau$-only integrability error $E_{\tau}$ accounts for a median 99.9\% of the actual $E$, and a linear regression of $E_{\tau\text{-only}}$ against $E_{\text{actual}}$ yields slope $= 0.969$ and $R^2 = 0.987$ [Fig.~\ref{fig:inverse}(d)], with $\kappa$ contributing only $\sim$0.1\% independently ($\Delta R^2 = 0.036$). This asymmetry reflects the physical rigidity of the C$_\alpha$--C$_\alpha$ virtual-bond angle ($\sigma_\kappa/\langle\kappa\rangle \approx 0.07$) compared to the conformational flexibility of the torsion angle ($\sigma_\tau/\langle\tau\rangle \approx 0.2$--1.1). For inverse design this implies that selecting amino acids with consistent $\tau$ preferences or residues that favor similar backbone torsion angles is more critical than matching a target curvature. The design strategy therefore reduces to two steps in which one first chooses a target $(\kappa_0, \tau_0)$ within the Ramachandran-reachable optimal interval, with a suggested range of $\kappa \in [1.45, 1.65]$\,rad and $\tau \in [0.70, 1.10]$\,rad that encompasses the Zone~A cluster with a margin for robustness. Second one selects a sequence whose residue-level $\tau$ values are as spatially uniform as possible, ensuring that the designed backbone remains within the integrable regime where the dispersion relation guarantees sub-angstrom reconstruction accuracy.

\begin{table}[b]
\caption{Inverse design parameters and uniform-helix reconstruction accuracy. The parameter space was scanned over $\kappa \in [1.2, 1.8]$\,rad, $\tau \in [0.3, 1.6]$\,rad on a $50 \times 50$ grid. The optimal design interval is defined by the Zone~A actual range with a suggested margin for robustness.}
\label{tab:inverse}
\begin{ruledtabular}
\begin{tabular}{lc}
\multicolumn{2}{c}{\textit{Parameter space}} \\
\colrule
Ramachandran-reachable fraction & 68.2\% (1706/2500) \\
Reachable $n_{\text{per turn}}$ range & 2.84--5.08 \\
\colrule
\multicolumn{2}{c}{\textit{Uniform-helix reconstruction (50 chains)}} \\
\colrule
Median RMSD (all chains) & 1.29\,\AA \\
RMSD $< 2$\,\AA\ success rate & 68\% (34/50) \\
Zone~A median RMSD & 0.91\,\AA \\
Zone~A success rate & 97\% (33/34) \\
\colrule
\multicolumn{2}{c}{\textit{$\tau$ dominance}} \\
\colrule
$E_{\tau\text{-only}}$ vs $E_{\text{actual}}$: slope & 0.969 \\
$E_{\tau\text{-only}}$ vs $E_{\text{actual}}$: $R^2$ & 0.987 \\
$\tau$ contribution (median) & 99.9\% \\
$\kappa$ independent contribution $\Delta R^2$ & 0.036 \\
\colrule
\multicolumn{2}{c}{\textit{Optimal design interval}} \\
\colrule
Zone~A actual: $\kappa$ & $[1.52,\,1.60]$\,rad \\
Zone~A actual: $\tau$ & $[0.84,\,1.02]$\,rad \\
Suggested design range: $\kappa$ & $[1.45,\,1.65]$\,rad \\
Suggested design range: $\tau$ & $[0.70,\,1.10]$\,rad \\
\colrule
\multicolumn{2}{c}{\textit{Canonical helix verification}} \\
\colrule
$\alpha$-helix: $n_{\text{per turn}}$ (predicted / exp.) & 3.62 / 3.6 \\
$3_{10}$-helix: $n_{\text{per turn}}$ (predicted / exp.) & 2.98 / 3.0 \\
$\pi$-helix: $n_{\text{per turn}}$ (predicted / exp.) & 4.20 / 4.4 \\
\end{tabular}
\end{ruledtabular}
\end{table}

\section{Discussion \& Conclusion}
\label{sec:discussion}

Integrable models have provided exact solutions and organizing principles across condensed-matter physics, nonlinear optics, and hydrodynamics, yet strictly global integrability is rarely realized in complex molecular systems. Crystalline solids harbor point defects and grain boundaries~\cite{irvine2013dislocation, quirk2024grain}, magnetic chains contain domain walls~\cite{mcroberts2024domain}, and optical fibers exhibit loss and dispersion management that break translational symmetry~\cite{leo2013nonlinear, roy2026study}. In each case the productive theoretical strategy is not to demand global integrability but to identify extended regions where the integrable description holds and to characterize the boundaries where it fails. The results presented here establish that the protein backbone admits precisely this treatment. The integrability error $E[n]$ is uncorrelated with chain length ($r = 0.13$, $p = 0.38$) and is spatially concentrated (Gini coefficient 0.55), confirming that it is a strictly local quantity determined by the residue-level geometric environment. A 37-residue chain such as LL-37 can simultaneously contain highly integrable helical segments and strongly non-integrable kink boundaries, and the $E[n]$ profile serves as the quantitative map that distinguishes them. The segmented prediction strategy that exploits this locality raises the success rate from 68\% to 88\% across the 50-chain dataset (Table~\ref{tab:prediction}), demonstrating that piecewise integrability is not merely a theoretical classification but a practical tool for analytic structure prediction.

The Jacobian analysis of the mapping between Ramachandran and Frenet angles elucidates a geometric mechanism underlying the chirality barrier that has long constrained the Hasimoto framework. The mapping preserves 93\% of the conformational information globally yet compresses the $\alpha_R$ basin into a narrow manifold in Frenet space where the condition number typically exceeds 30 and 68.4\% of grid points fall below $|J| = 0.1$. This geometric compression rather than a topological singularity at zero curvature renders the global inverse problem ill-posed by mapping distinct Ramachandran conformations to nearly indistinguishable Frenet parameters. The absence of a true singularity is supported by the physical bond constraints that maintain a minimum curvature of 0.58\,rad and implies that the coordinate transformation remains analytically valid despite its ill-conditioned nature. Within a local helical segment, however, this characteristic implies that the conformational diversity of the $\alpha_R$ basin collapses into a single near-uniform geometric state, so that the compression becomes benign and the Hasimoto field acts as a low-pass filter that strips high-frequency conformational noise while preserving the soliton envelope. A related consequence of this geometric dominance is that the DNLS potential becomes insensitive to sequence-specific bond lengths in the uniform-helix limit, rendering the system dependent only on the mean geometric parameters $(\langle\kappa\rangle, \langle\tau\rangle)$. The very property that prevents the Hasimoto representation from encoding chemical identity is what enables the analytic dispersion relation to predict backbone coordinates without sequence information. The same ill-conditioning also amplifies error accumulation in gradient-based refinement schemes, which the direct dispersion-relation approach circumvents by solving the geometry in Frenet space and performing coordinate reconstruction as a single non-iterative step.

The predominant role of torsion non-uniformity in driving integrability breaking stems from a distinct structural origin. The curvature $\kappa$ is geometrically rigid because the C$_\alpha$--C$_\alpha$ virtual-bond angle is primarily determined by the planar peptide unit which results in a relative variation of only $\sim$7\% ($\sigma_\kappa/\langle\kappa\rangle \approx 0.07$) across the dataset. The torsion $\tau$, by contrast, reflects the full dihedral freedom of the backbone modulated by side-chain packing, hydrogen bonding, and solvent exposure, with relative variations ($\sigma_\tau/\langle\tau\rangle$) spanning 0.2 to 1.1, an order of magnitude larger. Because $E[n]$ depends on $\tau$ through the site-to-site difference $\cos\tau[n] - \cos\tau[n{+}1]$, even modest torsion fluctuations produce non-negligible integrability breaking, whereas the curvature ratios $r^{\pm} \approx 1 \pm 0.07$ contribute only a near-unity multiplicative correction. Under the $\kappa$-uniform approximation, torsion accounts for a median 99.9\% of the total error with curvature contributing only $\Delta R^2 = 0.036$ independently (Table~\ref{tab:integrability}). This asymmetry translates directly into a design principle for inverse engineering of helical backbones. Within the Zone~A integrable regime (median reconstruction RMSD 0.91\,\AA, 97\% success), the design constraint reduces almost entirely to controlling torsion uniformity, and the practical strategy is to select amino acids with consistent backbone torsion preferences rather than to match a target curvature. The optimal design region defined by $\kappa \in [1.45, 1.65]$\,rad and $\tau \in [0.70, 1.10]$\,rad encompasses the $\alpha$-helical cluster with a margin for robustness and covers helices from $3_{10}$-like to $\pi$-like geometries within the Ramachandran-reachable region (Table~\ref{tab:inverse}).

The four-zone applicability classification (Table~\ref{tab:prediction}) provides explicit and falsifiable criteria for when the Hasimoto framework can and cannot predict structure. Zone~A ($\langle E\rangle < 0.10$, $\sigma_\tau < 0.40$\,rad) achieves 97\% success while Zone~D ($\langle E\rangle \geq 0.20$) achieves 0\% which constitutes a strict limit beyond which torsion heterogeneity is too pervasive for segmentation to resolve. This delineation distinguishes the present work from both data-driven methods such as AlphaFold~\cite{jumper2021highly} and ESMFold~\cite{lin2023evolutionary}, which achieve high accuracy through learned sequence-structure correlations but provide no analytic understanding of their applicability limits, and from the soliton-fitting approach of Niemi and collaborators~\cite{danielsson2010gauge, chernodub2010topological, molochkov2017gauge}, which uses the same DNLS framework in the descriptive direction to fit soliton parameters to observed $(\kappa,\tau)$ profiles. The dispersion-relation prediction is complementary to both. It requires no training data, no neural network, and no sequence information, and its accuracy within integrable islands (median RMSD 0.77\,\AA\ after segmentation) is competitive with the backbone accuracy of learned methods for helical segments. Several limitations naturally bound this complementarity. The entire analysis is confined to peptides with helical content exceeding 85\%, and the framework is not expected to extend to $\beta$-sheet or loop-dominated proteins where the $(\kappa,\tau)$ uniformity condition is not satisfied and where the 89.9\% $\beta$--PII curvature overlap (Table~\ref{tab:mapping}) precludes reliable conformational discrimination. The prediction requires an initial Frenet frame from the experimental structure and is therefore not a \textit{de novo} predictor but rather an analytic reconstruction and validation tool. Six chains (12\%) remain above 2\,\AA\ RMSD after segmentation, five of which share $\sigma_\tau > 0.80$\,rad, and the dataset includes sequence redundancy from multi-chain PDB entries (restricting to 32 unique structures preserves the zone-level conclusions with a segmented success rate of 84\%). An index-offset artifact at chain termini can also misattribute terminal torsion anomalies to interior residues, as observed for 3V1E\_A, suggesting that supplementing $E[n]$-based segmentation with direct $\tau$ sign-reversal detection would improve boundary robustness.

Two directions for future investigation emerge naturally from these findings. The segment-level integrability criterion could be combined with sequence-based helix propensity scales~\cite{chou1978empirical, pace1998helix} to predict from sequence alone which regions of a peptide are likely to fall within integrable islands, potentially enabling \textit{ab initio} analytic structure prediction for designed helical peptides without requiring experimental geometric input. The $\tau$-dominance finding further suggests that engineering torsion uniformity through careful residue selection may be a viable strategy for designing peptides with predictable helical geometry, with applications in antimicrobial peptide design where helical regularity correlates with membrane-disrupting activity~\cite{zelezetsky2006alpha, huang2014role}. More broadly, the $E[n]$ profile, previously established as a diagnostic of integrability breaking~\cite{wang2026Structural}, is here repurposed as a constructive tool that delineates integrable islands, guides segmented prediction, and defines the feasibility boundary for inverse design. The Hasimoto framework, within these quantitatively defined boundaries, functions as a precise local analytic instrument for the analysis and design of helical peptide geometry.

\appendix
\section{Per-chain prediction results}
\label{sec:appendix}

Table~\ref{tab:all_chains}lists the prediction results for all 50 helical peptide chains. Full-chain RMSD is computed from chain-averaged $(\langle\kappa\rangle, \langle\tau\rangle)$; segmented RMSD uses the $E > 0.10$ cutoff. Zone classification follows the criteria defined in Sec.~\ref{sec:results_prediction}.

\begin{table*}[htbp]
	\caption{Dispersion-relation prediction results for all 50 helical peptide chains. $N$: chain length (residues); $\langle E\rangle$: mean integrability error; $\sigma_\tau$: intra-chain torsion standard deviation (rad); Full: full-chain RMSD (\AA); Seg: segmented RMSD (\AA) calculated exclusively over the retained integrable islands; Islands: number of valid integrable segments ($\geq 4$ residues) retained after trimming; Coverage: percentage of total residues belonging to these islands. Chains are grouped by zone and sorted by $\langle E\rangle$ within each zone.}
	\label{tab:all_chains}
	\begin{ruledtabular}
		\begin{tabular}{llcccccccc}
			Chain & Zone & $N$ & Helix\% & $\langle E\rangle$ & $\sigma_\tau$ & Full (\AA) & Seg (\AA) & Islands & Cov.\,(\%) \\
			\colrule
			6XY0\_A& A & 30 & 100 & 0.014 & 0.042& 0.51 & 0.51 & 1 & 100 \\
			7NFG\_D& A & 30 & 97 & 0.016 & 0.048& 0.71 & 0.71 & 1 & 100 \\
			7Q1R\_A& A & 30 & 100 & 0.016 & 0.042& 0.64 & 0.64 & 1 & 100 \\
			8A3G\_B& A & 30 & 100 & 0.017 & 0.064& 0.49 & 0.49 & 1 & 100 \\
			8A3G\_A& A & 30 & 100 & 0.017 & 0.056& 0.50 & 0.50 & 1 & 100 \\
			7NFG\_E& A & 30 & 97 & 0.020 & 0.049& 0.78 & 0.78 & 1 & 100 \\
			4RWC\_A& A & 22 & 100 & 0.020 & 0.033& 0.23 & 0.23 & 1 & 100 \\
			7NFG\_C& A & 30 & 100 & 0.023 & 0.046& 0.78 & 0.78 & 1 & 100 \\
			5AL6\_A& A & 44 & 100 & 0.024 & 0.067& 1.02 & 1.02 & 1 & 100 \\
			7Q1Q\_A& A & 30 & 100 & 0.025 & 0.051& 0.52 & 0.52 & 1 & 100 \\
			7NFG\_B& A & 30 & 100 & 0.025 & 0.052& 0.87 & 0.87 & 1 & 100 \\
			3TQ2\_A& A & 35 & 97 & 0.025 & 0.128& 1.60 & 1.60 & 1 & 100 \\
			7NFG\_F& A & 30 & 97 & 0.026 & 0.059& 0.91 & 0.91 & 1 & 100 \\
			7NFG\_A& A & 30 & 100 & 0.027 & 0.052& 0.76 & 0.76 & 1 & 100 \\
			7MNK\_C& A & 30 & 93 & 0.027 & 0.061& 0.50 & 0.50 & 1 & 100 \\
			7Q1R\_B& A & 30 & 100 & 0.028 & 0.047& 0.62 & 0.62 & 1 & 100 \\
			7MNK\_A& A & 30 & 100 & 0.031 & 0.065& 0.54 & 0.54 & 1 & 100 \\
			6Q5O\_A& A & 30 & 97 & 0.039 & 0.163& 0.75 & 0.75 & 1 & 100 \\
			9HVN\_A& A & 22 & 100 & 0.039 & 0.062& 0.79 & 0.79 & 1 & 100 \\
			7Q1Q\_B& A & 30 & 100 & 0.042 & 0.120& 0.90 & 0.62 & 1 & 87 \\
			1ET1\_B& A & 34 & 97 & 0.042 & 0.199& 1.08 & 1.08 & 1 & 100 \\
			1ET1\_A& A & 34 & 97 & 0.045 & 0.182& 1.05 & 1.05 & 1 & 100 \\
			6YTU\_B& A & 33 & 97 & 0.045 & 0.125& 1.10 & 0.89 & 1 & 91 \\
			6YTU\_A& A & 34 & 97 & 0.046 & 0.218& 1.47 & 2.23 & 2 & 91 \\
			7BAS\_A& A & 29 & 97 & 0.054 & 0.260& 1.29 & 0.80 & 2 & 90 \\
			2WQ1\_A& A & 32 & 94 & 0.055 & 0.176& 1.48 & 0.85 & 1 & 91 \\
			7BAS\_D& A & 28 & 96 & 0.057 & 0.270& 1.31 & 0.55 & 2 & 89 \\
			7BAS\_E& A & 29 & 97 & 0.058 & 0.270& 1.30 & 0.79 & 2 & 90 \\
			7BAS\_B& A & 29 & 97 & 0.058 & 0.266& 1.29 & 1.15 & 2 & 90 \\
			7BAS\_C& A & 28 & 96 & 0.058 & 0.272& 1.29 & 0.26 & 2 & 89 \\
			2MLT\_A& A & 26 & 92 & 0.059 & 0.180& 2.50 & 0.85 & 2 & 92 \\
			2MAG\_A& A & 23 & 78 & 0.069 & 0.143& 0.96 & 0.60 & 2 & 65 \\
			2O6N\_A& A & 31 & 100 & 0.076 & 0.345& 1.94 & 0.94 & 2 & 81 \\
			3AZD\_B& A & 30 & 100 & 0.080 & 0.167& 0.93 & 0.58 & 3 & 90 \\
			\colrule
			5NMN\_A& B & 34 & 91 & 0.041 & 0.561& 1.65 & 0.74 & 1 & 88 \\
			6PHI\_A& B & 28 & 89 & 0.064 & 0.828& 3.00 & 0.29 & 1 & 75 \\
			5LHW\_A& B & 27 & 85 & 0.072 & 0.853& 2.76 & 0.84 & 2 & 89 \\
			4QK7\_A& B & 25 & 92 & 0.079 & 0.761& 3.20 & 0.33 & 1 & 88 \\
			4QKC\_A& B & 25 & 92 & 0.080 & 0.760& 3.20 & 0.31 & 1 & 88 \\
			3V1E\_A& B & 43 & 95 & 0.084 & 0.810& 16.72 & 11.16 & 3 & 93 \\
			6Q5Q\_A& B & 32 & 91 & 0.093 & 0.763& 2.81 & 0.80 & 1 & 88 \\
			\colrule
			3V1A\_A& C & 48 & 85 & 0.112 & 0.884& 19.38 & 1.89 & 2 & 75 \\
			1G2Y\_B& C & 29 & 86 & 0.153 & 1.350& 6.29 & 0.16 & 3 & 66 \\
			1G2Y\_C& C & 28 & 93 & 0.163 & 1.348& 7.21 & 3.99 & 3 & 71 \\
			1G2Y\_D& C & 29 & 86 & 0.164 & 1.295& 6.36 & 0.12 & 3 & 59 \\
			1G2Y\_A& C & 30 & 87 & 0.172 & 1.145& 6.75 & 0.54 & 3 & 67 \\
			2K6O\_A& C & 37 & 86 & 0.175 & 0.890& 4.92 & 0.48 & 3 & 57 \\
			1WY3\_A& C & 34 & 88 & 0.181 & 1.285& 7.67 & 4.15 & 4 & 71 \\
			\colrule
			2F4K\_A& D & 33 & 88 & 0.209 & 1.359& 7.98 & 4.70 & 3 & 55 \\
			7ARR\_C& D & 30 & 93 & 0.216 & 1.378& 10.02 & 3.18 & 2 & 43 \\
		\end{tabular}
	\end{ruledtabular}
\end{table*}

\bibliography{main}

\end{document}